\documentclass{article}
\usepackage[utf8]{inputenc}
\usepackage{authblk}
\usepackage{setspace}
\usepackage[margin=1.25in]{geometry}
\usepackage{graphicx}
\graphicspath{ {./figures/} }
\usepackage{subcaption}
\usepackage{amsmath}

\title{What is the Title of this Paper? Solving logic puzzles using algorithms.}

\author[1]{Ujaan Rakshit}
\author[2]{Nishchal Dwivedi}

\affil[1]{Delhi Public School International, Saket, New Delhi, India}
\affil[2]{Department of Basic Science and Humanities, SVKM’s NMIMS Mukesh Patel School of Technology Management \& Engineering, Mumbai, India}
\affil[1]{ujaanrakshit@gmail.com}
\affil[2]{nishchal.dwivedi@nmims.edu}

\date{}

\begin{document}

\maketitle

\begin{abstract}
This work delves into the realm of logic puzzles by focusing on the Knight and Knave problems popularized by Raymond Smullyan in his book series ``What is the Name of This Book?" The puzzles revolve around characters known as Knights (truth-tellers) and Knaves (liars), challenging solvers to determine the true identity of each person based on their statements. This work explores the utilization of Python algorithms to automate the process of solving these puzzles, offering a computational approach that enhances efficiency and accessibility. In this research we aim to develop a Python algorithm capable of parsing and analyzing the statements provided in the Knight and Knave puzzles. A logical reasoning framework is integrated within the algorithm to deduce the identities of the characters based on their statements. The algorithm processes the input statements, create a knowledge base, and make deductions following the rules of Knight and Knave logic. The developed algorithm is thoroughly tested on various instances of Knight and Knave puzzles, comparing its results to known solutions and manual approaches. We further expand the scope of the problem by introducing a Normal (who can sometimes lie and sometimes say the truth).

\end{abstract}

\section{Introduction}
Puzzles have long been a fascination of the human psyche \cite{rosenhouse2020games}\cite{ccakir2011bridging}, challenging our ability to reason and think critically\cite{michalewicz2011puzzle}. Among these puzzles, the problem of the Knights and Knaves has fascinated puzzle enthusiasts for decades. Written by the famous logician Raymond Smullyan in his 1978 book, ``What Is the Name of This Book? The Riddle of Dracula and Other Logical Puzzles"\cite{smullyan1986name}, these puzzles have characters known as Knights (always truth tellers) and Knaves (always liars), who provide the solver with puzzles that make sense. 

Python, renowned for its simplicity, readability, and extensive libraries, has emerged as a prominent language in various domains, including artificial intelligence, data analysis, and problem-solving. Its flexibility in handling logical operations, combined with its wide range of libraries, makes it a promising candidate for solving complex logical puzzles.

We are motivated to explore how to design algorithms to solve puzzles. Such an algorithm will require an expression of the puzzle in a logical form and the logic algebra should be used to solve these answers.


The Knight and Knave puzzles are based on a fictional island populated by inhabitants who can be classified as either Knights, Knaves or Normal people. Knights always tell the truth, Knaves always lie, while Normal people can do both at random. The objective is to identify the true identity of each person on the island based on their statements. These puzzles not only serve as an intellectual exercise but also as a practical application of logical reasoning, deductive inference, and problem-solving skills.

Traditionally, solving the Knight and Knave problems has relied heavily on manual reasoning and the use of truth tables, which can become increasingly complex as the number of characters and statements in a puzzle grows. With the public release of text-to-text AI models such as GPT-4 \cite{openai2023gpt4}, the question on the logical thinking ability of such models has been raised. This research seeks to leverage the power of Python to develop an algorithm capable of solving various instances of the Knight and Knave puzzles.

\section{Algorithm}
In this work, we have extended the assignment by the Harvard CS50 course\cite{MIT}. We studied their public code and generalised it\cite{github}.
The algorithm, coded in Python, consists of two separate files: logic.py and puzzle.py. The logic.py file is responsible for handling the conversion of logical statements presented in puzzle.py, which defines the specific puzzle being solved, into a format that can be processed by a computer. This algorithm is designed to solve problems that can be broken down using logical operations such as And, Or, Not, and Implication.

\subsection{And() Operator}
The And statement produces a positive output only if all the statements within the And() statement evaluate to true. In other words, if both X and Y are true, the And(X,Y) statement will evaluate to true; otherwise, it will evaluate to false. This can be visualized using a binary truth table.

\begin{displaymath}
\begin{array}{|c c|c|}
X & Y & X \land Y\\ 
\hline 
T & T & T\\
T & F & F\\
F & T & F\\
F & F & F\\
\end{array}
\end{displaymath}

\subsection{Or() Operator}
The Or statement produces a positive output if any or all of the statements within the Or() statement are true. For example, the Or(X,Y) statement will evaluate to true if either X or Y or both X and Y are true. It will only evaluate to false if both X and Y are false. This can be shown using a binary truth table. 

\begin{displaymath}
\begin{array}{|c c|c|}
X & Y & X \lor Y\\ 
\hline 
T & T & T\\
T & F & T\\
F & T & T\\
F & F & F\\
\end{array}
\end{displaymath}

\subsection{Not() Operator}
The Not statement negates the input it receives. For instance, the Not(X) statement will evaluate X as false, regardless of its initial truth value. This can be represented using a binary truth table. 

\begin{displaymath}
\begin{array}{|c|c|}
X & \bar{X}\\
\hline
T & F \\
F & T \\
\end{array}
\end{displaymath}

\subsection{Implication() Operator}
The Implication statement generates a positive output for any combination of truth values in the statements within the Implication() statement, except when the first statement is true and the second statement is false. For example, the Implication(X,Y) statement evaluates to true for all cases except when X is true and Y is false; in that particular scenario, it evaluates to false. This behavior can be demonstrated using a binary truth table.

\begin{displaymath}
\begin{array}{|c|c|c|}
X & Y & X \implies Y\\
\hline
F & F & T \\
F & T & T \\
T & F & F \\
T & T & T \\
\end{array}
\end{displaymath}

\section{Solving the Questions}
Puzzle.py defines the specific puzzle being solved, into a format that can be processed by a computer. This is done by first defining the problem as a set of logical operands and setting a set of basic rules.

\subsection{Symbols}
In order for the algorithm to determine the nature of each person, first a set of symbols needs to be determined. These symbols are

\begin{center}
XKnave, XKnight and XNormal
\end{center}

in which X can be replaced with any person as the problem determines. The algorithm uses these as flags set to either true and false, outputting the only one for which the problem is solved and hence have a true flag.

\subsection{Setting the Problem}
In order the solve the problem, first a problem needs to be determined. For example, a problem could be

\begin{center}
A says ``We are both Knaves."\\
B says nothing.
\end{center}

For this problem, the solution is, assuming neither are Normal, that A is a Knave and that B is a Knight.

\subsection{Defining the Problem}
Next, the problem has to be represented in logical statement form. The first rule to be defined is that both A and B are either a knight or knave, and that neither can be both. This can be defined as,

\begin{center}
And(Or(AKnave,AKnight), Not(And(AKnave,AKnight)))\\
And(Or(BKnave,BKnight), Not(And(BKnave,BKnight)))
\end{center}

Next, the statements made have to be represented using Implication() statements

\begin{center}
    Implication(AKnave,Not(And(AKnave,BKnave)))\\
    Implication(AKnight,(And(AKnave,BKnave)))
\end{center}

These statements state every case possible and outline what would happen for each case scenario of true and false symbols. Finally, all of the statements need to be combined by wrapping them in an And() logical operator

\begin{center}
    And(\\
    And(Or(AKnave,AKnight), Not(And(AKnave,AKnight))),\\
    And(Or(BKnave,BKnight), Not(And(BKnave,BKnight))),\\
    Implication(AKnave,Not(And(AKnave,BKnave))),\\
    Implication(AKnight,(And(AKnave,BKnave)))
)
\end{center}

Now, the algorithm can accept the input and run through the statements given to find the set of symbols for which all the statements hold true. This is only possible for AKnave and BKnight, giving a final output of

\begin{center}
    A is a Knave\\
    B is a Knight
\end{center}

\section{Results}
After rewriting a total of 15 different problems in such a manner as above, the following results were yielded.

\subsection{Regular Problems}
Regular problems are problems with only Knights and Knaves, no Normals, with only one possible solution. The algorithm was able to solve regular problems, such as the one above, with no specialities. For example, for the question

\begin{center}
    A says either ``I am a knight." or ``I am a knave.", but you don't know which.\\
    B says ``A said 'I am a knave'."\\
    B says ``C is a knave."\\
    C says ``A is a knight."
\end{center}

the algorithm correctly determines that

\begin{center}
    A is a Knight\\
    B is a Knave\\
    C is a Knight
\end{center}

\subsection{Indeterminate Problems}
Indeterminate problems are problems with only Knights and Knaves, no Normals, with multiple solutions. The algorithm was able output the determinate parts of indeterminate problems, that is to say, the parts that are true for every solution of the problem. For example, for the question

\begin{center}
    A says ``B is a knave."\\
    B says ``A and C are of the same type."
\end{center}

the possible solutions are that A is a knave, B is a knight and C is a knave, or that A is a knight, B is a knave and C is a knave. As can be seen, the only constant, or determinate part of the solution is that in both cases C is a knave. As such, the algorithm outputs

\begin{center}
    C is a Knave
\end{center}

\subsection{Normal Problems}
Normal problems are problems with have Knights, Knaves, and Normals. In Normal problems, it is assumed that, for a 3 person problem, one person is a knight, one person is a knave and the the last one is Normal. This is because, without this restriction, the Normal person could easily be substituted for a knight or a knave in case of a truth or a lie respectively. For example, for the question

\begin{center}
    A says ``I am Normal"\\
    B says ``That is true."\\
    C says ``I am not Normal."\\
    One is a knight, one is a knave, the other is Normal.
\end{center}

the same assumptions as before will not work. They have to be modified to account for the possibility of being Normal.

\begin{center}
        Or(\\
        And(And(AKnight, BKnave), CNormal),\\
        And(And(AKnight, CKnave), BNormal),\\
        And(And(BKnight, AKnave), CNormal),\\
        And(And(BKnight, CKnave), ANormal),\\
        And(And(CKnight, AKnave), BNormal),\\
        And(And(CKnight, BKnave), ANormal),
    )\\
    Not(And(ANormal,CNormal))\\
    Not(And(BNormal,CNormal))\\
    Not(And(CNormal,BNormal))
\end{center}

Doing so allows the algorithm to correctly determine that

\begin{center}
    A is a Knave\\
    B is Normal\\
    C is a Knight
\end{center}

\section{Discussion}
The utilization of Python algorithms in solving the Knight and Knave problems presents several advantages and implications worth discussing. This section explores the implications of employing computational approaches, the benefits of automation, and the limitations that may arise from such an approach.

The introduction of a Python algorithm to solve the Knight and Knave puzzles brings forth a new dimension to puzzle-solving methodologies. By automating the reasoning process, the algorithm provides a systematic and efficient approach that can handle complex scenarios with numerous characters and statements. This computational approach tries to substitute the need for manual reasoning with truth tables.

One significant advantage of the Python algorithm is its potential for broader accessibility. Traditional manual approaches to solving Knight and Knave puzzles often require a strong foundation in logical reasoning and a thorough understanding of truth tables. However, by implementing a computational algorithm, individuals with varying levels of expertise can engage with these puzzles. The algorithm acts as a guide, enabling users to analyze and solve the puzzles without being hindered by the complexities of manual reasoning.

Furthermore, the Python algorithm offers the advantage of speed and accuracy. As computers excel at processing large amounts of data and executing complex calculations, the algorithm can analyze multiple statements simultaneously and provide prompt and accurate solutions. This expedites the solving process, especially for puzzles with intricate scenarios, leading to increased efficiency and satisfaction for puzzle enthusiasts.

However, it is important to acknowledge certain limitations that may arise when using a computational approach. The algorithm relies heavily on the quality and accuracy of the input statements provided by the characters in the puzzles. Ambiguous or misleading statements can potentially lead to incorrect deductions or failed solutions. Therefore, the algorithm's performance is contingent upon the clarity and precision of the puzzle's design and the statements presented.

Moreover, the algorithm's success depends on the assumptions and constraints of the Knight and Knave logic. While the algorithm operates within the defined rules of the puzzles, it may face challenges when confronted with variations or extensions of the Knight and Knave problems that introduce additional elements or modified logic rules. Adapting the algorithm to handle such scenarios may require further modifications and adjustments.

\section*{Conclusion}
The work has shown how computational approaches can improve speed and accessibility in puzzle solving by introducing logical reasoning frameworks into the algorithm.
There are important ramifications from using computation to solve the Knight and Knave challenges.It offers an approach that is organised and effective and can deal with complex circumstances involving several characters and assertions, doing away with the necessity for manual reasoning and truth tables. 

The Python algorithm that was created has showed promise in parsing and analysing the statements given in the Knight and Knave riddles, correctly determining the identities of the characters based on their remarks. The algorithm's effectiveness has been tested on numerous problem examples, and the findings match up with known answers and manual methods. The algorithm serves as a guide, allowing users to analyse and solve the problems without being constrained by the difficulties of manual reasoning. As a result, anyone with diverse degrees of competence can participate in the puzzles.

However, it is also important to understand that a computational method has some potential drawbacks. The quality and accuracy of the input statements affect the algorithm's performance, and vague or deceptive statements may result in erroneous deductions. Furthermore, the method follows the established guidelines of the Knight and Knave logic, which may present difficulties when dealing with changes or modifications to the riddles.

In conclusion, the use of algorithms to solve the Knight and Knave riddles opens up new research opportunities for both puzzle fans and academicians. The automated method increases efficiency, accuracy, and accessibility, allowing a larger audience to enjoy these fascinating logic puzzles. However, the strength of the input statements and the constraints imposed by the Knight and Knave logic must be carefully taken into account. Future studies might examine how to modify the algorithm to tackle puzzles with more intricate variations or look into using various programming languages or methods to speed up the solution process. 

\bibliographystyle{unsrt}
\bibliography{bib}

\begin{thebibliography}{1}

\bibitem{rosenhouse2020games}
Jason Rosenhouse.
\newblock {\em Games for Your Mind: The History and Future of Logic Puzzles}.
\newblock Princeton University Press, 2020.

\bibitem{ccakir2011bridging}
Murat~Perit {\c{C}}ak{\i}r, Hasan Ayaz, Meltem {\.I}zzeto{\u{g}}lu, Patricia~A
  Shewokis, Kurtulu{\c{s}} {\.I}zzeto{\u{g}}lu, and Banu Onaral.
\newblock Bridging brain and educational sciences: An optical brain imaging
  study of visuospatial reasoning.
\newblock {\em Procedia-Social and Behavioral Sciences}, 29:300--309, 2011.

\bibitem{michalewicz2011puzzle}
Zbigniew Michalewicz, Nicholas Falkner, and Raja Sooriamurthi.
\newblock Puzzle-based learning: An introduction to critical thinking and
  problem solving.
\newblock {\em Decision line}, 42(5):6--9, 2011.

\bibitem{smullyan1986name}
Raymond Smullyan.
\newblock {\em What is the Name of this Book?}
\newblock Touchstone Books, 1986.

\bibitem{openai2023gpt4}
OpenAI.
\newblock Gpt-4 technical report, 2023.

\bibitem{MIT}
Cs50’s introduction to artificial intelligence with python.
\newblock https://cs50.harvard.edu/ai/2020/projects/1/knights/.

\bibitem{github}
https://github.com/KingSyclone/Knight-Knave-Normal.

\end{thebibliography}
\end{document}